\documentclass[5p,twocolumn,10pt]{elsarticle}
\usepackage{rotate}
\usepackage{lscape}
\usepackage{epsfig}
\usepackage{color}
\usepackage{amsmath}
\usepackage{subfiles}
\usepackage{subfig}
\usepackage{cite}

\newcommand{\bq}{\begin{equation}}
\newcommand{\eq}{\end{equation}}
\newcommand{\bqa}{\begin{eqnarray}}
\newcommand{\eqa}{\end{eqnarray}}
\newcommand{\baa}[1]{\begin{array}{#1}}
\newcommand{\eaa}{\end{array}}

\def\c2{\chi^2}

\def\gg{\gamma\gamma}

\def\gggg{\gamma\gamma\rightarrow\gamma\gamma}
\def\gggz{\gamma\gamma\rightarrow\gamma Z}
\def\ggzz{\gamma\gamma\rightarrow ZZ}

\newcounter{bla}

\journal{Computer Physics Communications}

\begin{document}
\begin{frontmatter}

\title{Monte-Carlo tool SANCphot for polarized $\gamma\gamma$ collision simulation}

\author[a,b]{Sergey~G.~Bondarenko}
\author[a]{Lidia~V.~Kalinovskaya}
\author[a]{Andrey~A.~Sapronov\corref{author}}

\cortext[author] {Corresponding author.\\\textit{E-mail address:} andrey.a.sapronov@gmail.com}
\address[a]{Dzhelepov Laboratory for Nuclear Problems, JINR, Joliot-Curie 6, RU-141980 Dubna, Russia}
\address[b]{Bogoliubov Laboratory of  Theoretical Physics, JINR, Joliot-Curie 6, RU-141980 Dubna, Russia}


\begin{abstract}
Our study of theoretical uncertainties for the four bosons processes at one-loop level including the case of the transverse polarization is presented. The calculations are based on helicity amplitudes approach for 4-boson SM interactions through a fermion and boson loops.
The computation  takes into account nonzero mass of loop particles. The obtained predictions are equally suitable for a wide range of energies and for arbitrary, including extreme, regions of the phase volume. Uncertainty estimates are received using the new Monte-Carlo tool SANCphot for $\gamma\gamma$ collision simulation with final states $\gamma\gamma$, $\gamma Z$, $ZZ$ adapted for polarized $\gamma$ beams.
\end{abstract}


\end{frontmatter}


{\bf PROGRAM SUMMARY}

\begin{small}
\noindent
{\em Program Title: sancphot-v1.01}                              \\
{\em Journal Reference:}                                      \\
{\em Catalogue identifier:}                                   \\
{\em Licensing provisions: none}                              \\
{\em Programming language: Fortran, C, C++}                   \\
{\em Computer: x86(-64) architecture}                                 \\
{\em Operating system: Linux}                                 \\
{\em RAM:} 2 Gbytes                                           \\
{\em Number of processors used: multiprocess}                 \\
{\em Classification:11.1, 11.6}                               \\
{\em External routines/libraries: }                     \\
{\em Subprograms used: Looptools~\citep{Hahn:1998yk}, Cuba~\citep{Hahn:2004fe}}              \\
{\em Nature of problem: Theoretical calculations at next-to-leading order in
perturbation theory allow to compute higher precision amplitudes for Standard
Model processes and decays, provided proper treatments of UV divergences and IR
singularities are performed.}\\
   \\
{\em Solution method: Numerical integration of the precomputed differential expressions for polarized photon-photon 
cross sections of four-boson interaction processes implemented as SANC modules~\citep{Andonov:2008ga}. }\\
   \\
{\em Restrictions: the list of processes is limited to photon-photon, photon-Z and Z-Z collisions}\\
   \\
{\em Running time: from hours to days depending on requested precision and kinematic conditions}\\
   \\

\end{small}

\section{Introduction \label{intro}}

The technology of photon-photon collider has been developed for decades~\citep{Ginzburg:1981ik, Ginzburg:1981vm, Ginzburg:1982yr}. As new TeV-scale $e+e-$ colliders being designed and rejected, the idea of $\gamma\gamma$ collider with photons generated from electron beams become more determined. For example, a concept of SLC-type $e^+e^-/\gamma\gamma$ facility at a Future Circular Collider is currently under development~\citep{Belusevic:2016odx}. In most cases the technology of high-energy photon generation uses the Compton scattering process of a laser light off the few-hundred GeV/TeV electron beams~\citep{Telnov:1989sd, Telnov:1995hc}. It is worth noting that such method does not require positrons, which are normally difficult to produce.

An overview of interesting physical processes which can be assessed using a photon collider is given in~\citep{Boos_2001}. The potential luminosity for such machine can reach up to 0.3 of luminosity of its ``parent'' lepton collider and in absolute numbers can be as high as $10^{34}\mathrm{cm}^{-2}\mathrm{s}^{-1}$. Such luminosity makes it possible to study such phenomena as SM and MSSM Higgs production, anomalous $WW$ interactions, photon structure functions.

The problems of light-by-light scattering and related 
radiative corrections were considered in a number of works~\citep{Karplus:1950zz,Jikia:1993tc,Bohm:1994sf, Gounaris:1998qk,Bern:2001dg}.
The article \citep{Gounaris:1999ux} looked into the process of $\gamma\gamma \to \gamma Z$.
A code for calculation of separate components ($\sigma_0, \sigma_{22},$ etc.) of ZZ production cross section in SM and MSSM models was presented in~\citep{Diakonidis:2006cv}. Two other works~\citep{Inan:2020gfu, Inan:2021pbx} investigate processes $\gggg$ and $\gggz$ with polarized incoming beams at CLIC in terms of anomalous couplings.

In this paper we present an original Monte-Carlo simulation tool for the photon-photon collisions incorporating the complete set of differential cross section expressed in helicity amplitudes convoluted with the Stocks parameters to account for the different polarization setups and two-boson final states. So far there was no publicly available program to calculate the full polarized photon-photon cross section at the 1-loop precision.

The program is based on the SANC library --- a collection of helicity amplitudes (HA) for the 4-boson interaction processes within the Standard Model. The SANCphot uses routines for differential cross sections produced by the SANC group~\citep{Andonov:2004hi, Bardin:2006sn}. The tool allows to calculate $\gamma\gamma$ cross section given such input parameters as initial electron energies and electron and photon polarization. Currently the processes with following final states are implemented: $\gamma\gamma$, $Z\gamma$, $ZZ$.

\section{Helicity amplitudes in SANC}

All the processes $\gamma \gamma \to \gamma \gamma$
$\gamma \gamma \to \gamma Z$
$\gamma \gamma \to ZZ$ can be treated as various cross channels 
of process  $f_1\bar{f}_1 bb\to 0$, and hence one-loop corrected scalar form factors, 
derived for this process, can be used for its cross channels also,
after an appropriate permutation of their arguments ($s,t,u$). This is not the case for
helicity amplitudes, however. They are different for all three channels and must be
calculated separately.
In our previous papers 
\citep{Bardin:2006sn, Bardin:2009gq, Bardin:2012mc, Bardin:2017qqw}
we consider  these processes in multi-channel approach, when all the particles participating in the process
are considered as incoming $\gamma\gamma bb \to 0$ ($b$, $\gamma$ or $Z$ boson). It was discussed all the one-loop diagrams, as well as their
corresponding amplitudes in terms of Lorenz expressions (based on the constructed basis) and scalar form factors.
Explicit expressions of the HA  are also given.

\section{Polarized photon beams}

As mentioned in Section~\ref{intro}, the photon beams are obtained using Compton scattering of laser photons from high energy electron beams. For example, we use the definitions for unpolarized and polarized $\gggg$ process cross sections introduced in~\citep{Gounaris:1999gh}, equation (10):
\small
\begin{align}
\begin{split}
{\frac{d\sigma}{d\tau d\cos\vartheta^*}}=&
{\frac{d\bar L_{\gamma\gamma}}{ d\tau}} \Bigg \{
{\frac{d\bar{\sigma}_0}{d\cos\vartheta^*}}
+\langle \xi_2 \xi_2^\prime \rangle
{\frac{d\bar{\sigma}_{22}}{d\cos\vartheta^*}} \\
& +[\langle\xi_3\rangle\cos2\phi+\langle\xi_3^ \prime\rangle\cos2\phi^\prime]
{\frac{d\bar{\sigma}_{3}}{d\cos\vartheta^*}} \\
&+\langle\xi_3 \xi_3^\prime\rangle[
{\frac{d\bar{\sigma}_{33}}{d\cos\vartheta^*}}
\cos2(\phi+\phi^\prime) \\
&+{\frac{d\bar{\sigma}^\prime_{33}}{d\cos\vartheta^*}}
\cos2(\phi- \phi^\prime)] \\
&+[\langle\xi_2 \xi_3^\prime\rangle\sin2 \phi^\prime-
\langle\xi_3 \xi_2^\prime\rangle\sin2\phi]
{\frac{d\bar{\sigma}_{23}}{d\cos\vartheta^*}} \Bigg \} \ \ , 
\label{sigpol}
\end{split}
\end{align}
\normalsize
where
\small
\begin{align}
{\frac{d\bar \sigma_0}{d\cos\vartheta^*}} & =
\left ({\frac{1}{128\pi\hat{s}}}\right )
\sum_{\lambda_3\lambda_4} 
 [|{\cal H}_{++\lambda_3\lambda_4}|^2
+|{\cal H}_{+-\lambda_3\lambda_4}|^2] ~ ,  \label{sig0}
\end{align}
\begin{align}
{\frac{d\bar{\sigma}_{22}}{d\cos\vartheta^*}} & =
\left ({\frac{1}{128\pi\hat{s}}}\right )\sum_{\lambda_3\lambda_4}
 [|{\cal H}_{++\lambda_3\lambda_4}|^2
-|{\cal H}_{+-\lambda_3\lambda_4}|^2]  \ , \label{sig22} \\
{\frac{d\bar{\sigma}_{3}}{d\cos\vartheta^*}} & =
\left ({\frac{-1}{64\pi\hat{s}}}\right ) \sum_{\lambda_3\lambda_4}
Re[{\cal H}_{++\lambda_3\lambda_4}{\cal H}^*_{-+\lambda_3\lambda_4}]  \ ,
\label{sig3} \\
{\frac{d\bar \sigma_{33}}{d\cos\vartheta^*}}& =
\left ({\frac{1}{128\pi\hat{s}}}\right ) \sum_{\lambda_3\lambda_4}
Re[{\cal H}_{+-\lambda_3\lambda_4}{\cal H}^*_{-+\lambda_3\lambda_4}] \ ,
\label{sig33} \\
{\frac{d\bar{\sigma}^\prime_{33}}{d\cos\vartheta^*}} & =
\left ({\frac{1}{128\pi\hat{s}}}\right ) \sum_{\lambda_3\lambda_4}
Re[{\cal H}_{++\lambda_3\lambda_4}{\cal H}^*_{--\lambda_3\lambda_4}] \  ,
\label{sig33prime} \\
{\frac{d\bar{\sigma}_{23}}{d\cos\vartheta^*}}& =
\left ({\frac{1}{64\pi\hat{s}}}\right ) \sum_{\lambda_3\lambda_4}
Im[{\cal H}_{++\lambda_3\lambda_4}{\cal H}^*_{+-\lambda_3\lambda_4}] \ ,
\label{sig23}
\end{align}
\normalsize
are expressed in terms of the $\gamma \gamma  \to bb$
helicity amplitudes, given in 
\citep{Bardin:2009gq,Bardin:2012mc, Bardin:2017qqw}. The expressions for the cross sections were compared symbolically and numerically with independent calculations in \citep{Bardin:2017qqw}.

The factor $d\bar{L}_{\gg}/d\tau$ denotes the photon-photon luminosity per unit $e^+e^-$ flux, considering the linear collider operating in $\gg$ mode. Further, $\vartheta^*$ is the $\gg$ collision products scattering angle in the incoming particles rest frame, and the $\tau\equiv s_{\gg}/s_{ee}$ -- fraction of electron beams energy transferred to the colliding photons.
The Stocks parameters $\xi_2, \xi_3$ are defined via the photon and electron polarization functions as described in~\citep{Ginzburg:1981vm},~\citep{Ginzburg:1982yr}.

It has to be admitted that the Compton scattering alone provides only approximate description of the incoming gamma spectrum. Nevertheless such calculation can be used for quantitative evaluation of model parameter dependencies.

\section{SANCphot numerical results}

The numerical validation was performed in the $\alpha(0)$ EW scheme using setup given below.
\begin{tabular}{ll}
$\alpha =1/137.035990996$,      & $G_{F} = 1.13024\times 10^{-5}\,\mathrm{GeV}^{-2}$,  \\
$M_{W}= 80.45149$\,GeV,           & $M_{Z}= 91.18670$\,GeV,            \\  
$M_{H}= 125$\,GeV,              &                                   \\
$m_e = 0.51099907$\,MeV,   & $m_\mu = 0.105658389$\,GeV,       \\
$m_\tau = 1.77705$\,GeV,        &                   \\
$m_u = 0.062$\,GeV,     & $m_d = 0.083$\,GeV,           \\
$m_c = 1.5$\,GeV,       & $m_s = 0.215$\,GeV,           \\
$m_t = 173.8$\,GeV,             & $m_b = 4.7 $\,GeV.           \\
\end{tabular} \\
Similarly to \citep{Gounaris:1998qk} the simulation of gamma-gamma collisions is approximated 
as if they were produced in the $ee$ collider framework. The initial electron energy was picked from 250~GeV, 
500~GeV, 1~TeV or 2~TeV, initial laser beam energy irradiating the electrons was set 
to 1.26120984~Ev. Such laser energy provides optimal 
conversion ratio for the backscattered photons with parameter $x_0 = 4.83$ (corresponding to electron beam energy 500~GeV).
The analytic expressions for the 4-boson interaction diverge at high scattering angle
making the numeric calculation unstable. Therefore the phase space is constrained by 
$|cos\theta| > 30^{\circ}$. Also the final state invariant mass for the processes $\gggg, \gggz, \ggzz$ was constrained to be above 20, 100 and 200~GeV correspondingly.

\onecolumn
\begin{figure}
  \begin{center}
\includegraphics[width=0.315\textwidth]{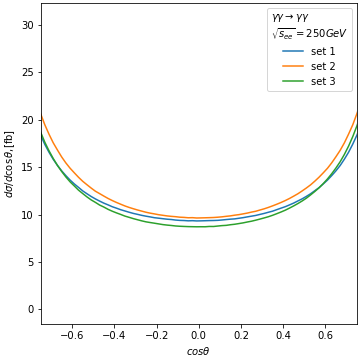}
\includegraphics[width=0.315\textwidth]{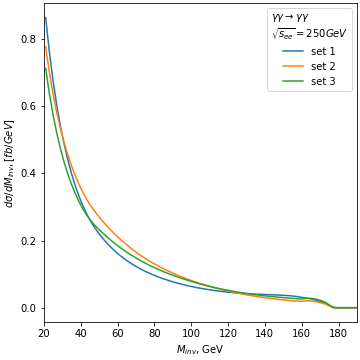}
\includegraphics[width=0.315\textwidth]{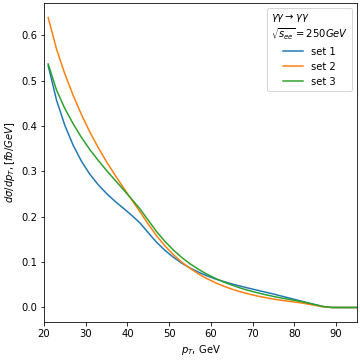} \\
\includegraphics[width=0.315\textwidth]{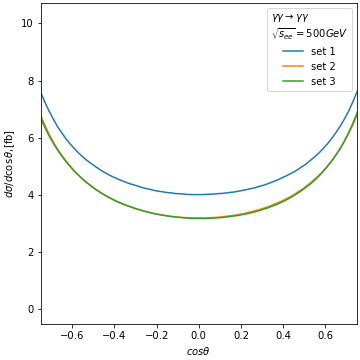}
\includegraphics[width=0.315\textwidth]{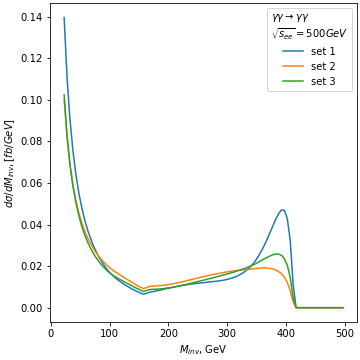}
\includegraphics[width=0.315\textwidth]{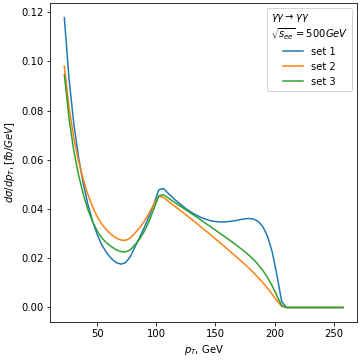} \\
\includegraphics[width=0.315\textwidth]{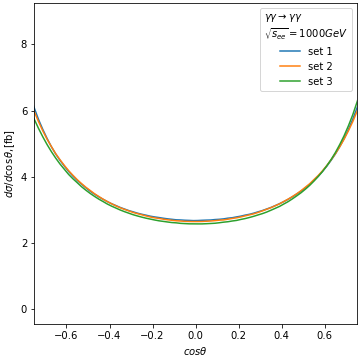}
\includegraphics[width=0.315\textwidth]{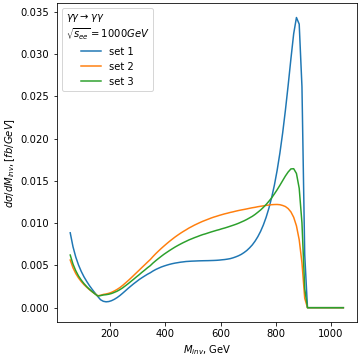}
\includegraphics[width=0.315\textwidth]{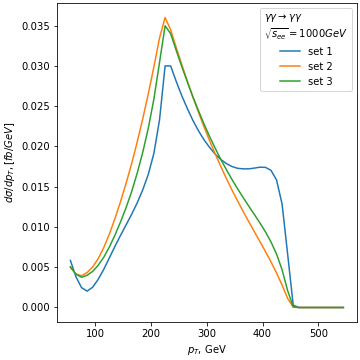} \\
\includegraphics[width=0.315\textwidth]{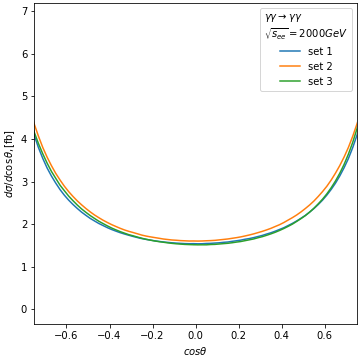}
\includegraphics[width=0.315\textwidth]{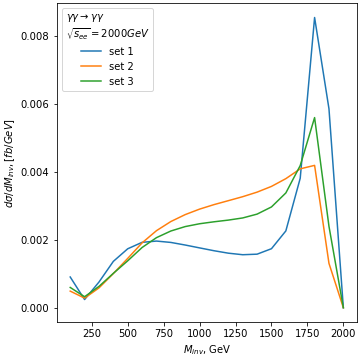}
\includegraphics[width=0.315\textwidth]{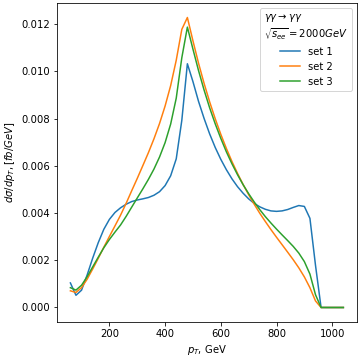} \\
  \end{center}
  \caption{Final state kinematic distributions for $\gggg$ process at different electron beam energies $\sqrt{s_{ee}} = 250, 500, 1000$ and $2000~\mathrm{GeV}$ 
  and various polarizations setups.\label{fig:kin_dist_gg}}
\end{figure}
\begin{figure}
  \begin{center}
\includegraphics[width=0.315\textwidth]{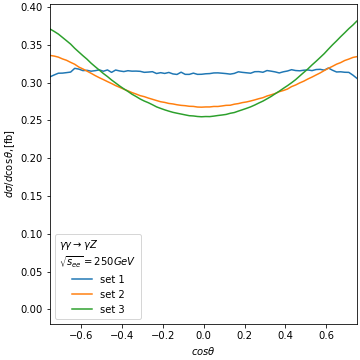}
\includegraphics[width=0.315\textwidth]{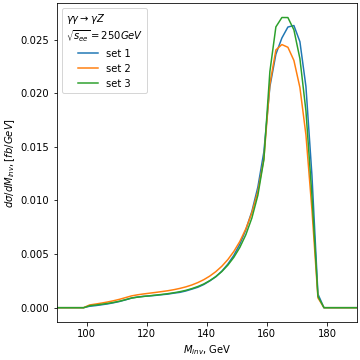}
\includegraphics[width=0.315\textwidth]{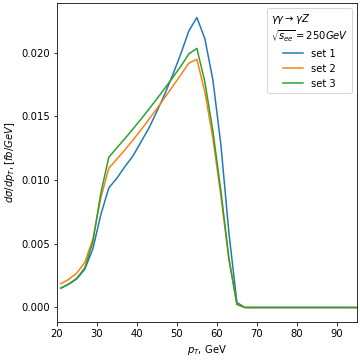} \\
\includegraphics[width=0.315\textwidth]{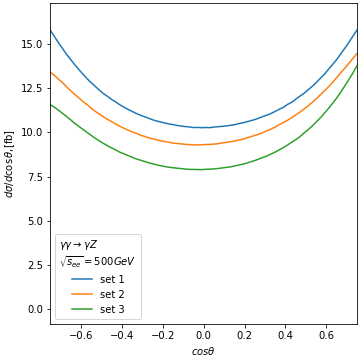}
\includegraphics[width=0.315\textwidth]{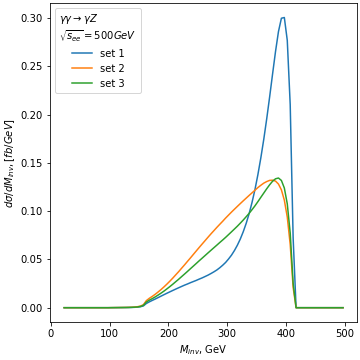}
\includegraphics[width=0.315\textwidth]{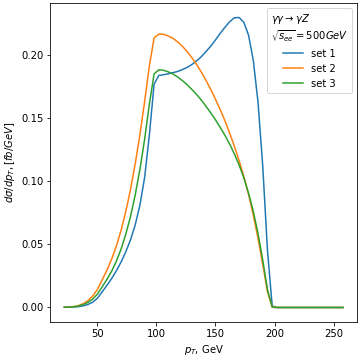} \\
\includegraphics[width=0.315\textwidth]{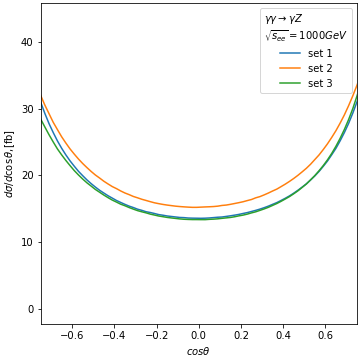}
\includegraphics[width=0.315\textwidth]{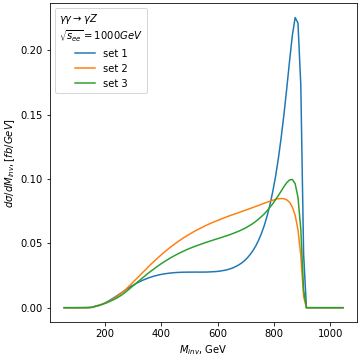}
\includegraphics[width=0.315\textwidth]{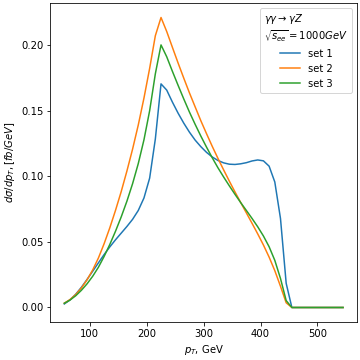} \\
\includegraphics[width=0.315\textwidth]{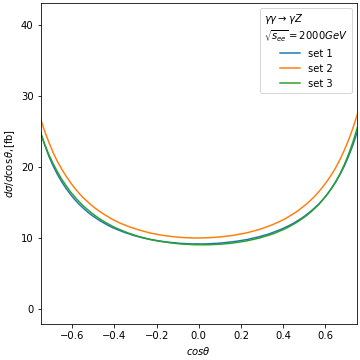}
\includegraphics[width=0.315\textwidth]{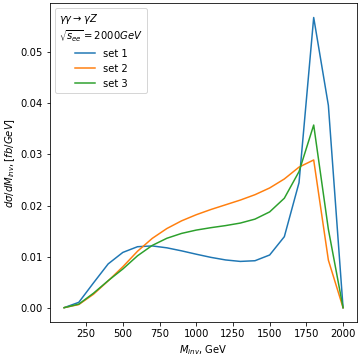}
\includegraphics[width=0.31\textwidth]{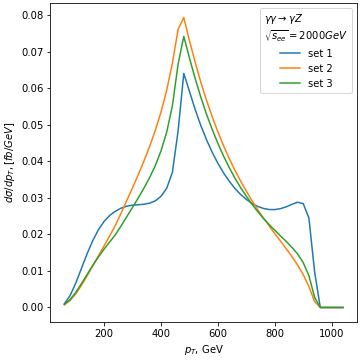} \\
  \end{center}
  \caption{Final state kinematic distributions for $\gggz$ process at different electron beam energies $\sqrt{s_{ee}} = 250, 500, 1000$ and $2000~\mathrm{GeV}$ 
  and various polarizations setups.\label{fig:kin_dist_gz}}
\end{figure}
\begin{figure}
  \begin{center}
\includegraphics[width=0.32\textwidth]{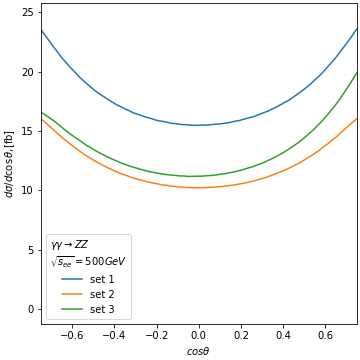}
\includegraphics[width=0.32\textwidth]{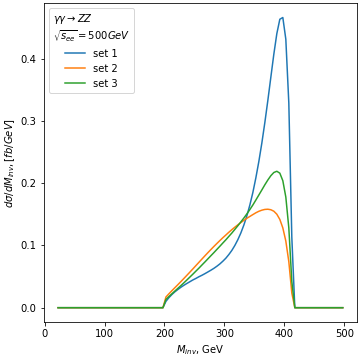}
\includegraphics[width=0.32\textwidth]{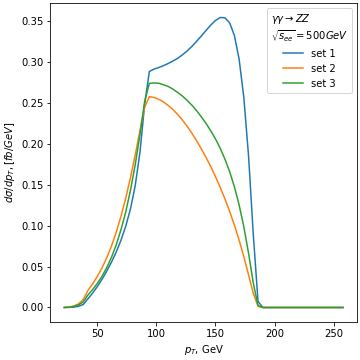} \\
\includegraphics[width=0.32\textwidth]{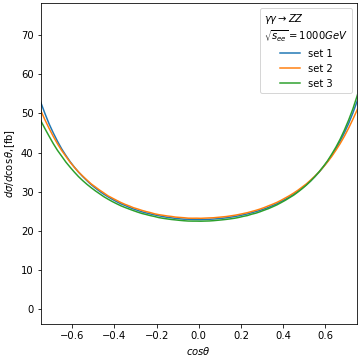}
\includegraphics[width=0.32\textwidth]{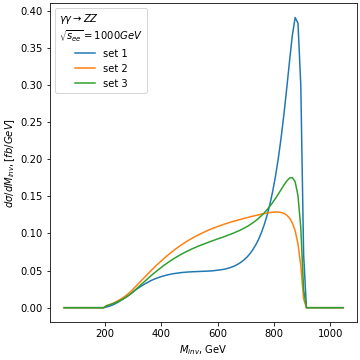}
\includegraphics[width=0.32\textwidth]{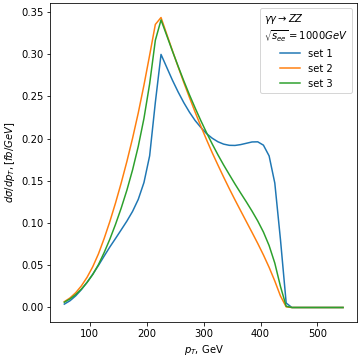} \\
\includegraphics[width=0.32\textwidth]{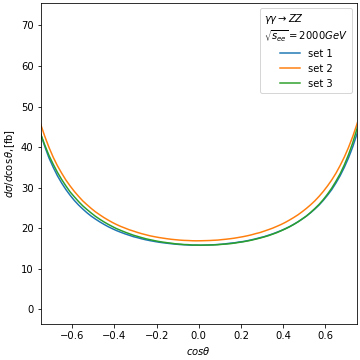}
\includegraphics[width=0.32\textwidth]{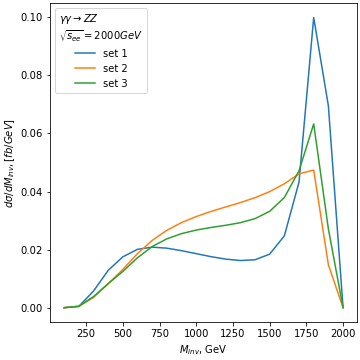}
\includegraphics[width=0.32\textwidth]{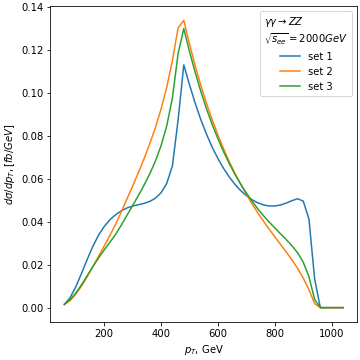} \\
  \end{center}
  \caption{Final state kinematic distributions for $\ggzz$ process at different electron beam energies $\sqrt{s_{ee}} = 500, 1000$ and $2000~\mathrm{GeV}$ 
  and various polarizations setups.\label{fig:kin_dist_zz}}
\end{figure}
\twocolumn
The cross section coefficients in Equation~\ref{sigpol} defined via the Stocks parameters 
$\xi_2, \xi_3$ that determine the normalized helicity density matrix of the backscattered 
photons. In order to probe dependence of the total cross section on the initial polarizations 
the numeric tests were conducted using the following three polarization combinations:
\begin{itemize}
\item set~1  : $P_e = P_e' = 0.8, P_{\gamma} = P_{\gamma}' = -1, P_t = P_t' = 0$
\item set~2  : $P_e = P_e' = 0,   P_{\gamma} = P_{\gamma}' = 0,  P_t = P_t' = 1, \phi=\pi/2$
\item set~3 : $P_e = 0.8, P_e' = 0, P_{\gamma} = -1, P_{\gamma}' = 0, P_t = 0, P_t' = 1, \phi=\pi/2$
\end{itemize}

Tables~\ref{tab_xs_gg}-\ref{tab_xs_zz} contain integrated cross sections for the three final states of the $\gamma\gamma$
collisions: $\gamma\gamma, \gamma Z, ZZ$. The cross section numbers for process $ZZ$ are not provided for $\sqrt{s_{ee}} = 250~\mathrm{GeV}$ as this energy is below the kinematic threshold for these final states.

\begin{table}
\centering
\begin{tabular}{|l|r|r|r|r|}
\hline
 $\sqrt{s_{ee}}$ & 250~GeV & 500~GeV & 1~TeV & 2~TeV \\
\hline
set 1  & 22.524(1)  & 9.4433(5)  & 6.9403(3)  & 4.3932(2) \\
set 2  & 24.426(1)  & 8.0456(4)  & 6.8598(3)  & 4.6406(2) \\
set 3  & 22.320(1)  & 8.0139(4)  & 6.7968(4)  & 4.4459(2) \\
\hline
\end{tabular}
  \caption{Integrated cross sections $\sigma(\gamma\gamma)$~[fb] for the process $\gamma\gamma \rightarrow \gamma\gamma$ in different polarization setups.\label{tab_xs_gg}}
\end{table}

\begin{table}
\centering
\begin{tabular}{|l|r|r|r|r|}
\hline
 $\sqrt{s_{ee}}$ & 250~GeV & 500~GeV & 1~TeV & 2~TeV \\
\hline
set 1  & 0.54590(3)  & 21.528(1)  & 35.157(2)  & 26.526(1) \\
set 2  & 0.51248(3)  & 19.374(1)  & 38.279(2)  & 28.887(1) \\
set 3  & 0.52999(3)  & 17.016(1)  & 34.528(2)  & 26.571(1) \\
\hline
\end{tabular}
  \caption{Integrated cross sections $\sigma(\gamma\gamma)$~[fb] for the process $\gamma\gamma \rightarrow \gamma Z$ in different polarization setups.\label{tab_xs_gz}}
\end{table}

\begin{table}
\centering
\begin{tabular}{|l|r|r|r|}
\hline
 $\sqrt{s_{ee}}$  & 500~GeV & 1~TeV & 2~TeV \\
\hline
set 1  &  32.335(2)  & 59.678(3)  & 45.900(2) \\
set 2  &  21.808(1)  & 58.945(3)  & 48.723(2) \\
set 3  &  24.319(1)  & 58.378(3)  & 46.381(2) \\
\hline
\end{tabular}
  \caption{Integrated cross sections $\sigma(\gamma\gamma)$~[fb] for the process $\gamma\gamma \rightarrow ZZ$ in different polarization setups.\label{tab_xs_zz}}
\end{table}


Figures~\ref{fig:kin_dist_gg}-\ref{fig:kin_dist_zz} show main kinematic distributions $cos\theta^*, M_{inv}$ and $p_T$ ($p_T(\gamma)$ for $\gggg$ and $\gggz$ processes or $p_T(Z)$ for $\ggzz$) 
for the three different final states calculated with various initial beams polarizations. 
Some asymmetry can be seen on the polar angle distributions due to the asymmetric polarization (set 3). 
On the invariant mass distributions a peak is observed near the kinematic threshold due to behaviour of 
the Stocks parameters.

\section{Summary \label{sec:summary}}
The presented \texttt{SANCphot} integrator is a Monte Carlo tool for evaluating
the $\gamma\gamma$ collisions with $\gamma\gamma, \gamma Z$ and $ZZ$ final states. 
The integrator is based on the
SANC framework modules and follows the same code organization as the \texttt{mcsanc} integrator~\citep{Bondarenko:2013nu, Arbuzov:2015yja}. 
Separate steps of the calculations were thoroughly cross checked against another sources including detailed symbolic and numeric comparisons and were shown to provide consistent results. 
The \texttt{SANCphot} uses advantage of multicore
implementation of the Cuba Monte Carlo integration library and supports
simple histogramming setup.

\section{Acknowledgements}
The authors are deeply thankful to the SANC group members for their help
and encouragement in developing the \texttt{SANCphot} integrator and writing
this paper. R.~Sadykov have significantly contributed to the results validation. 
The authors are also thankful to I.~Ginzburg for providing general recommendations on the approach and requirements for the \texttt{SANCphot} integrator development.

This research was funded by RFBR grant 20-02-00441.

\appendix

\section{SANCphot program \label{sec:sancphot-prog}}

\subsection{Installation}
There are two ways to install and run the \texttt{SANCphot} program with system install and 
docker container.

\subsubsection{System install}
After downloading and unpacking the tarball with
\texttt{SANCphot} from http://sanc.jinr.ru perform the following steps.
\begin{verbatim}
cd sancphot_vXX 
autoreconf --force --install
./configure
make
\end{verbatim}

If succeeded the executable will be created in \texttt{./src} directory. The program
is launched with a command from \texttt{./share}:
\begin{verbatim}
cd ./share
../src/sancphot [custom-input.cfg]
\end{verbatim}

\subsubsection{Docker container}
In order to avoid architecture and versioning problems, the distribution contains a 
Dockerfile which allows to build and run the code in a docker container. Provided
the docker is installed and running on the system, the \texttt{SANCphot} image can be 
built by issuing the command 
\begin{verbatim}
> docker build . -t sancphot:0.1
\end{verbatim}

Running the \texttt{SANCphot} in container:
\begin{verbatim}
docker run -e "CUBACORES=4"  \
    -v /full/path/to/share:/sancphot/share \
    sancphot:0.1 ../src/sancphot
\end{verbatim}

Upon completion an output file \texttt{sancphot}-\texttt{[run\_tag]}-\texttt{output.txt} with final results, run parameters and 
histograms will be created in the current directory.

\subsection{Configuration}
The \texttt{sancphot} program reads to configuration files upon start: \texttt{input.cfg} and \texttt{ewparam.cfg}.
The first file, \texttt{input.cfg}, contains general steering parameters for a run,
VEGAS parameters, kinematic cuts and histogram parameters organized in Fortran namelists.
\subsubsection{Process namelist}
\begin{description}
\item [processId] defines a process to calculate (011: $\gggg$, 012: $\gggz$, 022: $\ggzz$).
\item [run\_tag] is an arbitrary string value (\texttt{character*256}).
\item [sqs0] sets the electron beam energy in GeV at CMS frame (\texttt{double} \texttt{precision}).
\item [Elaser] sets the laser energy irradiating the electrons in eV (\texttt{double} \texttt{precision}).
\item [iflew(2)] flags controlling electroweak components of the NLO EW computations. See below. (\texttt{integer}).
\begin{description}
\item [iqed] = 0/1: corresponds to disabled/enabled QED corrections
\item [iew] = 0/1 corresponds to disabled/enabled weak corrections.
\end{description}
\end{description}

\subsubsection{Polarization namelist}
This namelist controls initial beams polarizations. For every parameter two values must be set: those of backward and forward beams.
\begin{description}
\item [laser\_Pl(2)] laser beam longitudinal polarization
\item [laser\_Pt(2)] laser beam transverse polarization
\item [laser\_phi(2)] rotation angle of the laser beam transverse polarization
\item [electron\_Pe(2)] electron beam polarization
\item [Z\_pol] Z-boson polarization (0: longitudinal, 1: transverse, 2: both)
\end{description}

\subsubsection{VegasPar namelist}
\texttt{VegasPar} regulates VEGAS parameters:
\begin{description}
\item [relAcc] relative accuracy limit on the calculated cross section (\texttt{double} \texttt{precision})
\item [absAcc] absolute accuracy limit on the cross section (\texttt{double} \texttt{precision})
\item [nstart] number of integrand evaluations performed at first iteration (\texttt{integer})
\item [nincrease] increment of integrand evaluations per iteration (\texttt{integer})
\item [nExplore] number of evaluations for grid exploration (\texttt{integer})
\item [maxEval] maximum number of evaluations (\texttt{integer})
\item [seed] random generator seed (\texttt{integer})
\item [flags] Cuba specific flags. See Cuba manual for more information~\citep{Hahn:2004fe} (\texttt{integer})
\end{description}

\subsubsection{KinCuts namelist}

This configuration section contains typical kinematic cuts for fiducial cross section calculation. Besides
original cuts, the user is free to implement his own in the \texttt{src/kin\_cuts.f}. The cut values
are set under corresponding name in the table in \texttt{input.cfg}

{\footnotesize
\begin{verbatim}
&KinCuts
  cutName       = 'm34',  'pt3', 'pt4',  ... 
  cutFlag       = 1,      1,     1,      ...
  cutLow        = 66d0,   20d0,  20d0,   ...
  cutUp         = 116d0,  7d3,   7d3,    ...
/
\end{verbatim}
}
The \texttt{cutName} fields are fixed, \texttt{cutFlag} - 0/1 on/off cuts.
By default seven kinematic cuts are foreseen. They are: invariant \texttt{m34}, transverse momenta \texttt{pt3, pt4},
and polar angle \texttt{theta3, theta4} of the final particles.

\subsubsection{Histograms with equidistant bins}

Namelist \texttt{FixedBinHist} allows to control histograms with fixed step. The histogram
parameters are defined in a table: under the histogram name, the user can set printing flag, 
min and max histogram values and a step. The binary printing flag controls if histogram is 
printed in the output (first bit) and if it's normalized to the bin width (second bit). 
For example to print unnormalized histograms set the flag to 1, to print normalized to 3.

{\footnotesize
\begin{verbatim}
&FixedBinHist
  fbh_name      = 'm34', 'pt3' 'pt4' ...
  fbh_flag      = 3,     3,    3,    ...
  fbh_low       = 66d0,  20d0, 20d0, ...
  fbh_up        = 116d0, 58d0, 58d0, ...
  fbh_step      = 2d0,   1d0,  1d0,  ...
/
\end{verbatim}
}

The most common parameters like invariant \texttt{m34}, transversal momenta \texttt{pt3, pt4} and polar angle cosines
\texttt{cth3, cth4} of the final particles are included by default. 

Any user defined histograms can be implemented in the \texttt{src/kin\_cuts.f} file. There 
the histogram names have no technical meaning, the requirement is that the order
of histograms in the input file must correspond to the histogram array \texttt{hist\_val} 
and the \texttt{nhist} variable set to the number of requested histograms.

\subsubsection{Histograms with variable bins}

Variable bin histograms are defined in namelist \texttt{VarBinHist}. The format is different
from \texttt{FixedBinHist}. The user have to set total number of variable step histograms
with \texttt{nvbh} parameter. Then follows definition of corresponding to \(i\)'th histogram name 
(\texttt{vbh\_name(i)}), printing flag (\texttt{vbh\_flag(i)}), 
number of bins (\texttt{vbh\_nbins(i)}) and array for the bin edges 
(\texttt{vbh\_bins(i,1:[nbins+1]}) as a space separated string of double precision numbers.

{\footnotesize
\begin{verbatim}
&VarBinHist
  nvbh          = 7,

  vbh_name(1)	= 'm34',
  vbh_flag(1)	= 3,
  vbh_nbins(1)	= 6,
  vbh_bins(1,1:7) = 50d0 55d0 60d0 70d0 100d0 200d0,

  vbh_name(2)	= 'pt3',
  vbh_flag(2)	= 0,
  vbh_nbins(2)	= 6,
  vbh_bins(2,1:7) = 50d0 55d0 60d0 70d0 100d0 200d0,

  vbh_name(3)	= 'pt4',
  vbh_flag(3)	= 3,
  vbh_nbins(3)	= 5,
  vbh_bins(3,1:10) = 25d0 30d0 50d0 70d0 90d0 110d0,

  ...

/
\end{verbatim}
}

Some common kinematic variables are included in variable bin histograms by default.

\subsection{Electroweak parameters}

Electroweak parameters are set in ewparams.cfg configuration file. It contains \texttt{\&EWPars} 
name list with the following fields.
\begin{description}
\item [gfscheme] flag defines electroweak scheme in which the calculation is performed (\texttt{integer}). For the $\gamma\gamma$ collision only value 0 corresponding to $\alpha(0)$ EW scheme is allowed.
\item [alpha, gf, conhc] - a list of constants and coefficients: 
  \(\alpha_{EM}\), \(G_{\mu}\), conversion constant (\texttt{double} \texttt{precision})
\item [mw, mz, mh] boson masses W, Z, Higgs (\texttt{double} \texttt{precision})
\item [wz, ww, wh, wtp] widths for bosons W, Z, Higgs and the top quark (\texttt{double} \texttt{precision}). These parameters are not used in the current integrator version. Reserved for future extension.
\item [men, mel, mmn, mmo, mtn, mta] leptonic masses \(\nu_e,e,\nu_{\mu}, \mu, \nu_{\tau}, \tau\)(\texttt{double} \texttt{precision})
\item [mdn, mup, mst, mch, mbt, mtp] quark masses \(u,d,s,c,b,t\)(\texttt{double} \texttt{precision})
\item [rmf1] is a 12 element array of fermion masses for technical purposes (\texttt{double} \texttt{precision})
\end{description}

Vegas grids have to be generated from scratch if these parameters are changed.

\subsection{Persistency}
Cuba library \citep{Hahn:2004fe} allows to save the VEGAS grid for further restoration.
The integration stops when requested accuracy is reached. The last version of the grid and 
histograms state are saved in \texttt{*.vgrid} and \texttt{*.hist} files. If the user desires
to improve precision he can set the new limits in \texttt{input.cfg}
and restart the program in the directory with saved grids. Care should be 
taken that all other configurations (energies, cuts, electroweak parameters) remains
the same. If the user wishes to change the histograms he can use 
old grids, but remove \texttt{*.hist} files and fill the histograms from scratch.
In this case the histogram errors will correspond to less precision than the total cross 
section.

The grid exploration stage is omitted if grid state file is found.

\subsection{Output}

The main output of the program will be written to
\texttt{sancphot}-\texttt{[run\_tag]}-\texttt{output.txt} file. The file contains
final table with cross sections from the components and the summed total cross
section. After the table follows the CPU usage, and a list of input parameters,
followed by a list of cuts applied in the process of integration. The remaining
of the output file is a section containing Histograms listing.  Each histogram
is output separately in a text file for plotting convenience. The file name
format is \texttt{[run\_tag]}\_\texttt{[hist\_name]}.


\bibliographystyle{utphys_spires}
\bibliography{main}

\end{document}